\documentclass[aps,prb,groupedaddress]{revtex4}
\usepackage{graphicx}

\begin{document}

%\preprint{}

\title{Absorbing processes in Richardson diffusion: analytical results}


\author{Simone Pigolotti}
%\thanks{}
\affiliation{The Niels Bohr Institut, Blegdamsvej 17, DK-2100
  Copenhagen, Denmark}
 \author{Mogens H. Jensen}
\affiliation{The Niels Bohr Institut, Blegdamsvej 17, DK-2100
  Copenhagen, Denmark} 
\author{Angelo Vulpiani}
\affiliation{Dipartimento di Fisica and INFN, Universit\`a di Roma "La
  Sapienza" P.le A. Moro 2, I-00185 Roma, Italy}


\date{\today}

\begin{abstract}
  We consider the recently addressed problem of a passive particle (a
  predator), being the center of a ``sphere of interception'' of
  radius $R$ and able to absorb other passive particles (the preys)
  entering into the sphere. Assuming that all the particles are
  advected by a turbulent flow and that, in particular, the Richardson
  equation properly describes the relative dispersion, we calculate an
  analytical expression for the flux into the sphere as a function of
  time, assuming an initial constant density of preys outside the
  sphere.  In the same framework, we show that the distribution of
  times of first passage into the sphere has a $t^{-5/2}$ power law
  tail, seen in contrast to the $t^{-3/2}$ appearing in standard $3D$
  diffusion. We also discuss the correction due to the integral length
  scale on the results in the stationary case.
\end{abstract}

%\keywords{}

\maketitle

%\section{}
%\subsection{}
%\subsubsection{}

The statistics of relative dispersion of scalars in a turbulent flow
is a topic of great theoretical and practical interest.  It is well
known, since the pioneering work of Richardson in 1926
\cite{richardson}, that relative particle dispersion in a turbulent
flow is superdiffusive, in particular that the mean-square separation
between the particles grows with the third power of the time. Richardson's
explanation for this phenomenon was that, as the distance between
particles increases, the effective diffusion should also increase
since larger and larger eddies are involved. He was able to measure
that the effective diffusion is proportional to $r^{4/3}$, where $r$ is
the particle separation, and suggested a $3D$ diffusion equation for
the interparticle distance (now known as the Richardson equation) with
a diffusion coefficient growing like $r^{4/3}$ that indeed reproduces
the correct superdiffusive behavior.  Apart from practical
implication, Richardson's result may be considered {\it a posteriori}
as the background for Kolmogorov's development of the idea of
universality in turbulent flows \cite{k1941}.

Indeed, within Kolmogorov theory, the $4/3$ exponent is simply
obtainable by a dimensional argument: we will assume in the following
that both the initial and the final separation are much smaller than
the integral length scale $L$ (the typical scale of the largest
eddies) and much larger than the Kolmogorov scale
$\eta=(\nu^3/\epsilon)^{1/4}$, where $\nu$ is the viscosity and
$\epsilon$ is the mean dissipation rate of turbulent kinetic energy.
Then, the effective diffusion can depend only on the interparticle
distance $r$ and the strength of the field, which is proportional to
$r^{1/3}$.  The result is $D_{eff}=C\epsilon^{1/3}r^{4/3}$, where $C$
is a universal constant.  Plugging this coefficient into a $3D$
diffusion equation in spherical coordinates, one obtains:
\begin{equation}\label{richardsoneq}
\partial_t f(r,t)= \nabla_r [D_{eff}(r) \nabla_r f(r,t)]  =C \epsilon^{1/3} r^{-2}\partial_r [r^{10/3} \partial_r f(r,t)].
\end{equation}
By solving the equation with the intial condition $r(t=0)=0$, one can
easily show that:
\begin{equation}\label{separation}
\langle r^2(t)\rangle \propto \epsilon\ t^3.
\end{equation}

Nowadays, we have a far better comprehension of turbulent phenomena,
and we know that Kolmogorov theory is only approximatively valid
because of intermittency. Still, the predictions of
Eq.(\ref{separation}) are found to be practically unaffected by
intermittency corrections\cite{boffetta1,boffetta2}: the robustness of
this result is related to the fact that the mean dissipation rate
$\epsilon$ appears linearly into the equation.

Actually, many phenomena of interest in chemical and biological
sciences involving diffusion fall in the class of the absorption
problems. A typical example are the so-called reaction rate problems
in chemistry \cite{rice}: consider a particle $A$ surrounded by an
uniform distribution of diffusing particles $B$, and able to react
instantaneously with them, $A+B\rightarrow A$, when they fall inside a
sphere of radius $R$, centered in the position of $A$.  The problem is
then to calculate the flux inside the sphere as a function of time. In
$3D$ Brownian diffusion, the time dependent flux writes $\Phi(t)=4\pi
R\mathcal{D}[1+R/\sqrt{\pi \mathcal{D}t}]$ where $\mathcal{D}$ is the
diffusion coefficient; this classical result was firstly derived by
Smoluchowski\cite{smoluchowski}. The same problem in the $2D$ case,
formulated as the mean area spanned by a disc undergoing Brownian
motion in a time $t$, was solved by Kolmogorov and Leontovitch some
years later \cite{kolmogorovarea}.  Another problem one could consider
is the distribution of the first passage times: if there are few
reactants, then it may be interesting to know the probability for a
reaction to occur in a time $t$.

The reaction rate problem was recently addressed for the description
of plankton predator-prey dynamics in turbulent flows
\cite{osborn,predatorprey}. Here the particle $A$ (the predator)
belongs to a plankton species feeding on the preys $B$, and one would
like to calculate the amount of preys consumed per unit time.  The
biological assumptions underlying this model are essentially that the
predator is able to eat instantaneously every prey in its ``sphere of
interception'' of radius $R$, and that both the predator and the preys
are passive particles, i.e. their velocities are exactly the velocity
of the fluid at their positions.

On the physical side, it is tempting to face these absorption problems
in turbulent fluids using a simple approach based on the Richardson
equation; still, it is not trivial to assess whether this effective
description of the evolution of the interparticle distance is valid
when one considers problems that are more subtle than growth of the
the mean square separation with time. We just point out here that some
of these problems (like the first passage times problem) are known to
be related to backward diffusion, and recently \cite{sawford} it has
been shown that non-Gaussian tails and temporal correlations in the
Eulerian flow may imply relevant differences between the forward and
backward diffusion, like a different, non-universal value of the
constant $C$. Still, it is encouraging that some predictions based on
the Richardson equation for absorbing processes have been tested
experimentally \cite{predatorprey,occupationsphere} and numerically
\cite{pecseli}.

In this Letter, we show how these problems can be analytically tackled
assuming the validity of the Richardson equation. Our main results are
the analytical expression of the time-dependent flux, that is the
solution of the reaction rate problem for the Richardson diffusion,
and the calculation of the exponent of the power-law tails of the
first passage time distribution. We also discuss, in the stationary
case, the corrections due to the presence of an integral length scale
$L$ above which the diffusion coefficient does not depend on $r$
anymore.

Eq. (\ref{richardsoneq}) is casted into a standard Fokker-Planck form
\cite{gardiner} by defining the radial density $p(r,t)=r^2f(r,t)$:
\begin{equation}\label{eq:FP}
  \partial_t p(r,t)=C \epsilon^{1/3}\partial_r\left[-\frac{10}{3}r^{1/3}p(r,t)+\partial_r r^{4/3}p(r,t)\right].
\end{equation}
In order to simplify the expression, we make the substitution $r^{1/3}=x$
\begin{equation}\label{mainFP}
\partial_t g(x,t)=C\epsilon^{1/3} \partial_x\left[-\frac{8}{x}g(x,t)+\partial_x g(x,t)\right]
\end{equation}
Notice that this equation is effectively the same as the equation for a
$9$-dimensional diffusion in radial coordinate. The subject of our
study is obtained by taking the adjoint of the operator in the r.h.s
of Eq.(\ref{mainFP}), that is the backward Kolmogorov equation
corresponding to the Richardson equation:
\begin{equation}\label{backward}
\partial_t f(x,t)=C\epsilon^{1/3}\left[\frac{8}{x}\partial_x f(x,t) +\partial^2_x f(x,t)\right].
\end{equation}

Eq.(\ref{backward}) constitutes the appropriate equation to describe
absorbing processes; indeed, by imposing absorbing boundary conditions
on Eq.  (\ref{backward}), $f(R^{1/3},t)=f(\infty,t)=0\ \forall t$, the
function $f(x,t)$ represents the probability of not leaving the
interval $[R^{1/3},\infty)$ in a time $t$, i.e. of not being absorbed
into a sphere of radius $R$, for a particle starting at position $x$.
It can also be shown \cite{redner} that, calling $\theta(x,t)$ the
radial concentration of reacting chemicals (or preys, in the language
of reference\cite{predatorprey}), and imposing slightly modified
boundary conditions $\theta(R^{1/3},t)=0$, $\theta(\infty,t)=\rho$
where $\rho$ is the initial (constant) density of preys, then the
function $\theta(x,t)$ still evolves according to Eq.(\ref{backward}).

By introducing the Laplace transform of $f(x,t)$
with respect to time, $\tilde{f}(x,t)=\int_0^\infty dte^{-st}f(x,t)$, 
Eq. (\ref{backward}) becomes:
\begin{equation}\label{mainlaplace}
\frac{8}{x}\partial_x \tilde{f}(x,s) +\partial^2_x \tilde{f}(x,s)-s\tilde{f}(x,s)=0.
\end{equation}
If we find a solution $h(x)$ of Eq. (\ref{mainlaplace}) for $s=1$ which
is real and positive for $x>1$, then it is easy to show that
\begin{equation}
\tilde{f}(x,s)=\frac{1}{s}\frac{h(x\sqrt{s})}{h(\sqrt{s})}
\end{equation}
is the complete solution to Eq. (\ref{mainlaplace}) with the 
boundary conditions corresponding to the absorption
problem, while the reaction rate problem is solved 
by $\tilde{\theta}(x,s)=\rho[1/s-\tilde{f}(x,s)]$.

The function $h(x)$ is, up to a multiplicative constant, found to be 
\begin{equation}
h(x)=x^{7/2} K_{7/2}(x) = \sqrt{\frac{2}{\pi}}{x^{7}}(15+15x+6x^2+x^3)e^{-x}
\end{equation}
where $K$ is the modified Bessel function of the second kind \cite{abramovitz}, and we have
written its expression in terms of simple functions. 

Let us now move to the calculation of the time-dependent flux into the sphere:
\begin{equation}
\tilde{\phi}_R(S)=-D(r)\frac{\partial \theta(r,s)}{\partial r}|_{r=R}=-\frac{7C\epsilon^{1/3}\rho R^{1/3}}{3s}+\rho R^{2/3}\sqrt{\frac{C\epsilon^{1/3}}{s}} 
\frac{K_{5/2}(R^{1/3}\sqrt{s/(C\epsilon^{1/3})})}{K_{7/2}(R^{1/3}\sqrt{s/(C\epsilon^{1/3})})}
\end{equation}
and the integrated flux $\tilde{\Phi}_R(s)=4\pi R^2\tilde{\phi}_R(s)$ equals 
\begin{equation}\label{totalfluxlaplace}
\tilde{\Phi}_R(s)=\frac{28\pi C\epsilon^{1/3}\rho R^{7/3}}{3s}-4\pi \rho R^{8/3}\sqrt{\frac{C\epsilon^{1/3}}{s}}\frac{K_{5/2}(R^{1/3} 
\sqrt{s/(C\epsilon^{1/3})})}{K_{7/2}(R^{1/3}\sqrt{s/(C\epsilon^{1/3})})}
\end{equation}
Next we apply the inverted Laplace transform. Notice that the first
term in Eq. (\ref{totalfluxlaplace}) corresponds to the $t\rightarrow
\infty$ steady state flux: $\Phi_\infty=28\pi C\epsilon^{1/3}R^{7/3}/3$.
We invert the Laplace transform of the following function:
\begin{equation}
  \label{eq:invertfunc}
 \tilde{f}(s)=\frac{1}{\sqrt{s}}\frac{K_{5/2}(\sqrt{s})}{K_{7/2}(\sqrt{s})}=\frac{3+3\sqrt{s}+s}{15+15\sqrt{s}+12s+s^{3/2}}
\end{equation}
and the inverse Laplace transform is given by the Bromwich integral
\begin{equation}
f(t)=\frac{1}{2\pi i}\int_{\alpha-i\infty}^{\alpha+i\infty} ds\ \tilde{f}(s)\ e^{ts} ds
\end{equation}
where $\alpha$ is an arbitrary real number such that all the
singularities of the integrated function lie on the left of the
contour of integration. To avoid problems with branch cuts, we make
the substitution $\sqrt{s}=z$
\begin{equation}
f(t)=\frac{1}{2\pi i}\int_{\Gamma_1} dz \frac{6z+6z^2+2z^3}{15+15z+12z^2+z^3} e^{tz^2} dz
\end{equation}
where the contour path is $\Gamma_1$ in the picture. We evaluate
the integral over the closed contour of the figure: it is possible to
calculate the poles of the integrand, they have negative real parts
and are thus outside the integration contour. It is also possible to
show that the contribution of the arcs vanishes according to Jordan's
Lemma.
\begin{figure}[htbe]
\includegraphics[width=8cm]{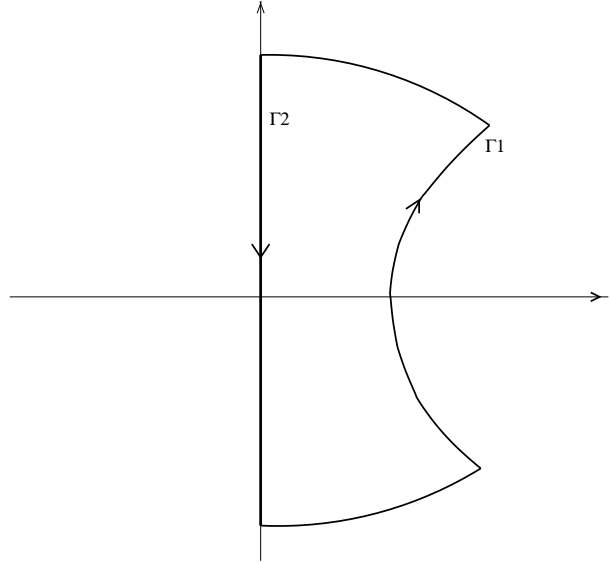}
\caption{Integration contour in the complex plane.}
\end{figure}
Then we have:
\begin{eqnarray}\label{funcexp}
\frac{1}{2\pi i}\int_{\Gamma_1} dz\ 2z\tilde{f}(z^2) e^{tz^2} dz=-\frac{1}{2\pi i}\int_{\Gamma_2} dz\ 2z\tilde{f}(z^2) e^{tz^2} =\nonumber\\
=-\frac{1}{2\pi}\int_{-\infty}^{+\infty}dy\ \frac{2y^4(18+y^2)}{225-135y^2+114y^4+y^6}e^{-ty^2}dy 
\end{eqnarray}
where the variable is $y=iz$. The ratio of polynomials in
(\ref{funcexp}) is obtained as the real part of the function
$2z\tilde{f}(z^2)$ calculated on the imaginary axis and its imaginary
part vanishes when integrated since it is an odd function of the
variable $y$.

\begin{figure}[htb]
\includegraphics[width=8cm]{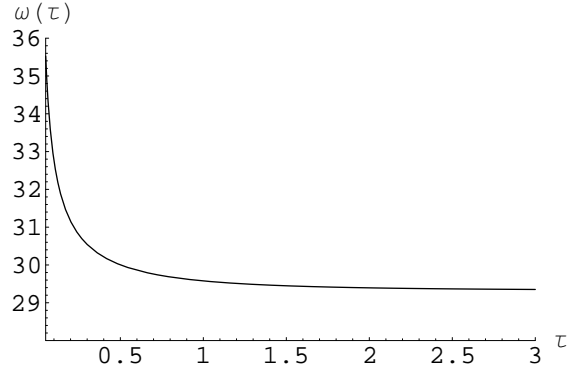}
\caption{\label{scaling}Plot of the scaling function $\omega(\tau)$}
\end{figure}

Going back to our problem, the time solution is found to be:
\begin{equation}
\Phi_R(t)=\frac{28\pi C\epsilon^{1/3}\rho R^{7/3}}{3}+2C\epsilon^{1/3}\rho R^{7/3} 
\int_{-\infty}^{+\infty}dy\ \frac{2y^4(18+y^2)}{225-135y^2+114y^4+y^6}e^{-\frac{C\epsilon^{1/3}t}{R^{2/3}}y^2}dy 
\end{equation}
and, as it is expected, can be cast into the scaling form
$\Phi_R(t)=C\rho\epsilon^{1/3}R^{7/3}\omega(C\epsilon^{1/3}t/R^{2/3})$.
The shape of the scaling function $\omega(\tau)$, where the
non-dimensional time $\tau$ is equal to $C\epsilon^{1/3}t/R^{2/3}$, is
shown in Fig.(\ref{scaling}).
Notice also that the flux diverges like $\tau^{-1/2}$ when
$\tau\rightarrow0$.

\begin{figure}[htb]
\includegraphics[width=8cm]{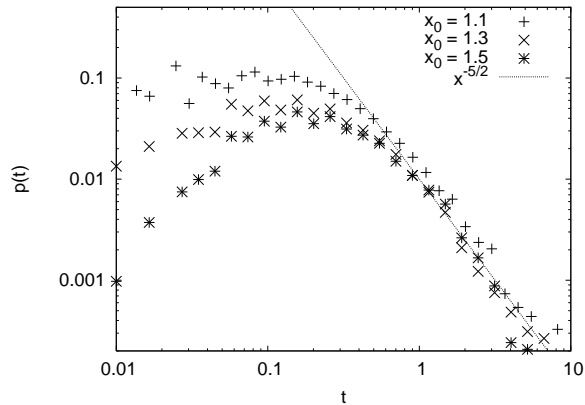}

\caption{\label{firstpassfig}Probability density function of the first passage times. 
  The statistics for different initial conditions, indicated in the
  legend, is collected from $5\cdot10^6$ realization of the Langevin
  equation corresponding to eq. (\ref{mainFP}).  A $t^{-5/2}$ power
  law is also shown for reference.}
\end{figure}

We move now to the problem of estimating the probability for a
particle to be absorbed in a time $t$, given the starting distance
$x$. We already know the Laplace transform of this probability
$\tilde{f}(x,s)$, which is the solution of Eq.(\ref{mainlaplace}). Our
goal is to calculate the inverse transform:
\begin{equation}
f(x,t)=\frac{1}{2\pi i}\int_{\alpha-i\infty}^{\alpha+i\infty}\tilde{f}(x,s)e^{ts}ds
\end{equation}
In order to do that, we call $z=\sqrt{s}$ and perform the integral
over the same contour as in  the previous problem. In this case, we have
to avoid a singularity in the origin which gives a time-independent
contribution corresponding to the finite probability of being absorbed
for $t\rightarrow\infty$.  The time behavior is then given by the
principal part of the following integral:
\begin{equation}
p(t)\propto P\int_{-\infty}^{+\infty}dy\ \frac{k(x,y)}{x^7y(225+45y^2+6y^4+y^6)}e^{-ty^2} dy
\end{equation}
with:
\begin{eqnarray}
k(x,y)=6y(75(x-1))-5(-1+3x(2+x(5x-2)))y^2+2x^2(15x-1)y^4\cos(yx-y)+\nonumber\\
+6(75-15(x-2)(2x-1)y^2-x(5+3x(25x-4))y^4+5x^3y^6)\sin(y-xy)\quad .
\end{eqnarray}
The first non-zero term in the expansion is proportional to
$t^{-3/2}$, meaning that the first-passage time have a $t^{-5/2}$
power law tail. In Fig. \ref{firstpassfig}, we compare simulations of
the first passage process with this theoretical prediction, by a
simple numerical integration of the Langevin equation corresponding to
the Fokker Planck equation (\ref{mainFP}).

Now, we want to show that the effect of the integral length scale is
just to introduce a small correction to the stationary flux, by
solving the stationary problem $\nabla_r [D_{eff}(r) \nabla_r f(r)]
=0$, where $D_{eff}(R)$ is equal to $C\epsilon r^{4/3}$ for $r<L$,
being $L$ is the integral length scale defined in the beginning, and
it is constant, $D_{eff}(r)=C\epsilon R^{4/3}$, for $r>L$. In $3D$,
the equation becomes $r^{-2}D_{eff}(r)\partial_r f(r)=const$, where
the value of the constant is determined by the boundary conditions
$f(\infty)=\rho$ and $f(R)=0$:
\begin{equation}
const=\rho\left(\int_R^\infty \frac{dr}{r^2 D_{eff}(R)} \right)^{-1}=\frac{7\rho}{C\epsilon^{1/3}\left(3R^{-7/3}+4L^{-7/3}\right)}
\end{equation}
so that the stationary flux becomes:
\begin{equation}
\Phi_R(t)=\frac{28\pi C\epsilon^{1/3}\rho R^{7/3}}{3+4(R/L)^{7/3}}
\end{equation}
As one could expect, the effect of the integral length scale is to
{\em decrease} the flux, and the correction can be safely neglected
for realistic values of the parameters ($R$ is supposed to be much
smaller than $L$).

To conclude, by assuming the validity of a description {\it \`a la}
Richardson for backward relative diffusion in turbulent flows, we show
how simple problems like the calculation of the flux into a moving
absorbing sphere and the first passage times distribution are
analytically solvable. We also discuss the effect of the integral
length scale on these results.  Our solution could provide a basis for
building up more realistic and detailed models of absorption phenomena
in turbulent flows.

\begin{acknowledgments}
  We are grateful Hans P\'ecseli for introducing us to the
  predator-prey scenario for passively advected particles and for many
  discussions and suggestions during the project. We acknowledment
  partial support from the EU Network {\it Stirring and Mixing}.
\end{acknowledgments}

\end{document}